\documentclass[12pt, titlepage]{article}
\usepackage{graphicx}
\usepackage{amsmath,amssymb}
\usepackage{amsthm}
\usepackage{algorithmic}
\usepackage{verbatim}
\usepackage{algorithm,algorithmic}
\def\x{{\mathbf x}}
\def\e{{\mathbf e}}
\def\u{{\mathbf u}}
\def\a{{\mathbf a}}
\def\b{{\mathbf b}}
\def\y{{\mathbf y}}

\begin{document}

\title{TECHNICAL REPORT:\\ TWO OBSERVATIONS ON PROBABILITY DISTRIBUTION SYMMETRIES FOR RANDOMLY-PROJECTED DATA }
%

\date{May 18, 2012}
\author{Hanchao Qi \and Shannon Hughes \\ Department of Electrical, Computer, and Energy Engineering\\ University of Colorado at Boulder
       }

\newtheorem{thm}{Theorem}
\newtheorem{lem}{Lemma}
\newenvironment{prf}[1][Proof]{\begin{trivlist}
\item[\hskip \labelsep {\bfseries #1}]}{\end{trivlist}}
\newtheorem{cor}{Corollary}

\maketitle

In this technical report, we will make two observations concerning symmetries of the probability distribution resulting from projection of a piece of $p$-dimensional data onto a random $m$-dimensional subspace of $\mathbb{R}^p$, where $m < p$.  In particular, we shall observe that such distributions are unchanged by reflection across the original data vector and by rotation about the original data vector.  

To start, let us introduce some notation.  Suppose that ${\bf x}$ is the original piece of data in $\mathbb{R}^p$.  We randomly generate $m$ vectors $\e_1, \ldots, \e_m \in \mathbb{R}^p$ from the Gaussian distribution with mean $0$ and covariance the identity matrix $I_{p\times p}$.  We then form the matrix $E \in \mathbb{R}^{p \times m}$ with columns $\e_1, \ldots, \e_m$.  The projection of $\x$ onto the $m$-dimensional subspace of $\mathbb{R}^p$ spanned by $\e_1, \ldots, \e_m$ is then given by $P\x$ where $P = E(E^TE)^{-1}E^T$.  

We can now prove the following two lemmas about the distribution of the random vector $P\x$.  The first shows that the distribution of $P\x$ is unchanged when reflected across $\x$.  The second shows that this distribution is unchanged when rotated about the axis of $\x$.  

\begin{lem}[Symmetry of the distribution of $P\x$ under reflection across $\x$]
\label{lem1}
Suppose $\x$ is a fixed point in $\mathbb{R}^p$ and let $P\x$ be a random vector with $P$ as defined above. Define the reflection operator $R_{\x}$ as
\begin{equation}
R_{\x}(\y) = \y + 2\left(\langle \y,{\hat \x}\rangle {\hat \x}-\y\right) = 2\langle \y,{\hat \x}\rangle {\hat \x}-\y
\end{equation}
where ${\hat \x} = {\x \over \|{\x}\|}$. Then the distribution of $P\x$ is the same as the distribution of $R_\x(P\x)$.
\end{lem}

\begin{prf}
For every realization $\e_1^0, \ldots , \e_m^0$ of the random variables $\e_1, \ldots , \e_m$, there is an equally likely realization $R_\x(\e_1^0), \ldots , R_\x(\e_m^0)$. This can be easily seen from the fact that the Gaussian distribution $\mathcal{N}(0,I_{p \times p})$ is symmetric across any line through the origin of $\mathbb{R}^p$. \\

We will show that if we define $P_{(\e_1^0, \ldots , \e_m^0)}(\x)$ as the projection of $\x$ onto the subspace spanned by $\e_1^0, \ldots , \e_m^0$. Then
\begin{equation}
P_{(R_{\x}(\e_1^0), \ldots , R_\x(\e_m^0))}(\x) = R_\x(P_{(\e_1^0, \ldots ,\e_m^0)}(\x)).
\end{equation}
That is, the projection of $\x$ onto the reflected vectors $R_{\x}(\e_1^0), \ldots , R_\x(\e_m^0)$ is the reflection of that onto the original random vectors $\e_1^0, \ldots, \e_m^0$.

To show this, we first observe three properties of the operator $R_{\x}$:

\begin{itemize}
	\item {\bf Property 1:} $R_\x$ is a linear operator:
\begin{eqnarray*}
R_\x(\alpha \a+\beta \b) &=&  2\langle\alpha \a+\beta \b,{\hat \x}\rangle{\hat \x}-\alpha \a-\beta \b \\
&=& \alpha\left(2\langle \a,{\hat \x}\rangle{\hat \x}-\a\right)+\beta\left(2\langle \b,{\hat \x}\rangle{\hat \x}-\b\right)\\
&=&\alpha R_\x(\a)+\beta R_\x(\b)
\end{eqnarray*}

\item {\bf Property 2:} The operator $R_\x$ preserves inner products (and hence norms as well):
\begin{eqnarray*}
\langle R_\x(\a),R_\x(\b)\rangle &=&  \langle 2\langle \a,{\hat \x}\rangle{\hat \x}-\a,2\langle \b,{\hat \x}\rangle{\hat \x}-\b\rangle \\
&=& 4\langle \a,{\hat \x}\rangle \langle \b,{\hat \x}\rangle - 4\langle \a,{\hat \x}\rangle \langle \b,{\hat \x}\rangle+\langle \a,\b \rangle\\
&=&\langle \a,\b \rangle
\end{eqnarray*}

\item {\bf Property 3:} For any orthonormal $\u_1, \ldots ,\u_k$ and any $\b$, the projection of $R_\x(\b)$ onto $R_\x(\u_1), \ldots , R_\x(\u_k)$ is the reflection of that of $\b$ onto $\u_1, \ldots ,\u_k$: \\
Using the first two properties above, and the fact that $R_\x(\u_1), \ldots, R_\x(\u_k)$ must be orthonormal by Property 2, we have:
\begin{eqnarray*}
P_{(R_\x(\u_1), \ldots, R_\x(\u_k))}\left(R_\x(\b)\right) &=& \sum_{j=1}^k \langle R_\x(\b), R_\x(\u_j)\rangle R_{\x}(\u_j) \\ &=& \sum_{j=1}^k \langle \b, \u_j \rangle R_\x(\u_j) \\ &=& R_\x\left(\sum_{j=1}^k \langle \b,\u_j\rangle \u_j\right) \\ &=& R_\x\left(P_{(\u_1, \ldots, \u_k)}(\b)\right)
\end{eqnarray*} 
\end{itemize}

Using the above three properties, we can easily see that if we perform Gram-Schmidt orthogonalization on $\e_1^0, \ldots, \e_m^0$ to obtain orthonormalized vectors $\u_1, \ldots , \u_m$, then performing Gram-Schmidt orthogonalization on $R_\x(\e_1^0), \ldots , R_\x(\e_m^0)$ must result in $R_\x(\u_1), \ldots , R_\x(\u_m)$.  To see this, we note that Gram-Schmidt involves two alternating steps: (i) we subtract from the currently selected vector its orthogonal projection onto those orthonormal vectors already obtained and (ii) we scale the resulting vector by 1 over its norm.  Suppose that we start with the two sets of vectors $\e_1^0, \ldots, \e_m^0$ and $R_\x(\e_1^0), \ldots , R_\x(\e_m^0)$.  We note that the second set are initially the reflections of the first set.   If we run the steps of Gram-Schmidt on the two sets of vectors simultaneously, then each step of Gram-Schmidt preserves the property that the second set of vectors are the reflections of the first set.  In the case of step (i), the orthogonal projections that we subtract off from the second set are reflections by Property 3 above of those we subtract off from the corresponding vector in the first set.  Then, the linearity of $R_\x$ (Property 1 above) guarantees that the resulting difference vector in the second set is a reflection of that obtained for the first set.  In the case of step (ii), the norms we divide by are equal (Property 2 above).

Hence, we find that using Property 3 above and the fact that $R_\x(\x) = \x$, we have that: 
\begin{equation*}
 P_{(R_\x(\e_1^0), \ldots ,R_x(\e_m^0))}(\x) = P_{(R_\x(\u_1), \ldots , R_\x(\u_m))}(\x)) 
 = R_\x(P_{(\u_1, \ldots , \u_m)}(\x))
 = R_\x(P_{(\e_1^0, \ldots , \e_m^0)}(\x))
\end{equation*}

Finally, since for every realization $\e_1^0, \ldots, \e_m^0$ of the random variables $\e_1, \ldots , \e_m$, resulting in the projection $P\x = P_{(\e_1^0, \ldots , \e_m^0)}(\x)$, there is an equally likely realization $R_\x(\e_1^0), \ldots, R_\x(\e_m^0)$, resulting in the projection $P_{(R_\x(\e_1^0), \ldots ,R_x(\e_m^0))}(\x) = R_\x(P_{(\e_1^0, \ldots , \e_m^0)}(\x)) = R_\x(P\x)$, we have that the probability distribution $f$ of $P\x$ satisfies
\begin{equation*}
f(P\x) \leq f\left(R_\x(P\x)\right). 
\end{equation*}

Similarly, since for every realization $R_\x(\e_1^0), \ldots, R_\x(\e_m^0)$, resulting in the projection $R_\x(P\x)$, there is an equally likely realization $R_\x(R_\x(\e_1^0)), \ldots, R_\x(R_\x(\e_m^0)) = \e_1^0, \ldots, \e_m^0$, resulting in the projection $P\x$, we have that
\begin{equation*}
	f\left(R_\x(P\x)\right) \leq f(P\x).
\end{equation*}

These inequalities show that
\begin{equation*}
f(P\x) = f\left(R_\x(P\x)\right).
\end{equation*}

This proves Lemma 1. \qed \\[6pt]
\end{prf}

\begin{lem}[Symmetry of the distribution of $P\x$ under rotation about $\x$]
Suppose $\x$ is a fixed point in $\mathbb{R}^p$ and let $P$ be as defined above. Let $V \in \mathbb{R}^{p \times p}$ be an orthogonal matrix with first column ${\hat \x} = {\x \over \|{\x}\|}$ and let
\begin{equation*}
Q_{\x} = V\left( \begin{array}{cc}  1 & 0_{1 \times (p-1)} \\  0_{(p-1) \times 1} & Q \end{array} \right)V^T
\end{equation*}
where $Q$ is in the special orthogonal group $SO_{p-1}$, so that $Q_{\x}$ represents an arbitrary rotation of $\mathbb{R}^p$ about $\x$.  Then the distribution of $P\x$ is the same as the distribution of $Q_{\x}(P\x)$.
\end{lem}

\begin{prf}
The proof follows the exact same structure as that of Lemma 1.

Similarly, we note that for every realization $\e_1^0, \ldots , \e_m^0$ of the random variables $\e_1, \ldots , \e_m$, there is an equally likely realization $Q_{\x}\e_1^0, \ldots ,Q_{\x}\e_m^0$, since the Gaussian distribution is rotationally symmetric. \\

Then we would like to show that if we define $P_{(\e_1^0, \ldots , \e_m^0)}(\x)$ as the projection of $\x$ onto the subspace spanned by $\e_1^0, \ldots , \e_m^0$, then
\begin{equation}
\label{eq:PQ}
P_{(Q_{\x}\e_1^0, \ldots , Q_{\x}\e_m^0)}(\x) = Q_{\x}P_{(\e_1^0, \ldots ,\e_m^0)}(\x).
\end{equation}
That is, the projection of $\x$ onto the rotated vectors $Q_{\x}\e_1^0, \ldots , Q_\x\e_m^0$ is the rotation of that onto the original random vectors $\e_1^0, \ldots, \e_m^0$.

As before, to prove this, we first note that $Q_{\x}$ is a linear operator.  $Q_{\x}$ also preserves inner products and norms (i.e.~$\langle Q_{\x}(\a), Q_{\x}(\b)\rangle = \langle \a,\b\rangle$ for all $\a$ and $\b$) since $V$ is an orthogonal matrix and $Q$ is in the special orthogonal group $SO_{p-1}$.

Using these two properties, we can show that for any vector $\b$ and any orthonormal set $\u_1, \ldots, \u_k$, we have that 
\begin{eqnarray*}
P_{(Q_\x \u_1, \ldots, Q_\x \u_k)}\left(Q_\x \b\right) &=& \sum_{j=1}^k \langle Q_\x \b, Q_\x \u_j\rangle Q_{\x} \u_j \\ &=& \sum_{j=1}^k \langle \b, \u_j \rangle Q_\x \u_j \\ &=& Q_\x\left(\sum_{j=1}^k \langle \b,\u_j\rangle \u_j\right) \\ &=& Q_\x P_{(\u_1, \ldots, \u_k)}(\b)
\end{eqnarray*}

The same argument as before can be used with the above three properties to show if $\u_1, \ldots, \u_m$ is the result of Gram-Schmidt orthogonalization on the vectors $\e_1^0, \ldots, \e_m^0$, then $Q_\x \u_1, \ldots, Q_\x \u_m$ must be the result of the Gram-Schmidt orthogonalization on $Q_\x \e_1^0, \ldots, Q_\x \e_m^0$.  

Finally, using the above and the fact that $Q_\x \x =\x$, we see that
 \begin{equation*}
 P_{(Q_{\x}(\e_1^0), \ldots , Q_{\x}(\e_m^0))}(\x) = P_{(Q_\x(\u_1), \ldots, Q_\x(\u_m))} (\x) = Q_\x P_{(\u_1, \ldots, \u_m)}(\x) = Q_{\x}(P_{(\e_1^0, \ldots , \e_m^0)}(\x))
 \end{equation*}

Since for every realization $\e_1^0, \ldots, \e_m^0$ of the random variables, $\e_1, \ldots, \e_m$, resulting in the projection $P_{(\e_1^0, \ldots , \e_m^0)}(\x)$, there is an equally likely realization $Q_{\x}(\e_1^0), \ldots , Q_{\x}(\e_m^0)$, resulting in the projection $P_{(Q_{\x}(\e_1^0), \ldots , Q_{\x}(\e_m^0))}(\x) = Q_{\x}P_{(\e_1^0, \ldots , \e_m^0)}(\x) = Q_\x P\x$, we see that the probability distribution $f$ of $P\x$ satisfies 
\begin{equation*}
f(P\x) \leq f(Q_{\x}P\x). 
\end{equation*} 
Moreover, noting that the matrix 
\begin{equation*}
Q_{\x}^{-1} = V\left( \begin{array}{cc}  1 & 0_{1 \times (p-1)} \\  0_{(p-1) \times 1} & Q^{-1} \end{array} \right)V^T
\end{equation*}
has the same properties as $Q_\x$, we can see that for every realization $Q_\x \e_1^0, \ldots, Q_\x \e_m^0$ of $\e_1, \ldots, \e_m$, resulting in the projection $Q_\x P\x$, there is an equally likely realization $Q_\x^{-1}Q_\x \e_1^0, \ldots, Q_\x^{-1} Q_\x \e_m^0 = \e_1^0, \ldots, \e_m^0$, resulting in the projection $P\x$.   We therefore also have
\begin{equation*}
	f(Q_{\x}P\x) \leq f(P\x).
\end{equation*}
  
These inequalities show that 
\begin{equation*}
f(P\x) = f(Q_{\x}P\x).
\end{equation*} 

 This proves Lemma 2. \qed
 \end{prf}
\end{document}